\documentclass{article}

\usepackage{arxiv}

\usepackage[utf8]{inputenc} 
\usepackage[T1]{fontenc}    
\usepackage{hyperref}       
\usepackage{url}            
\usepackage{booktabs}       
\usepackage{amsfonts}       
\usepackage{nicefrac}       
\usepackage{microtype}      
\usepackage{lipsum}
\usepackage{upgreek}
\usepackage{graphicx}
\DeclareUnicodeCharacter{3B1}{\ensuremath{\alpha}}
\DeclareUnicodeCharacter{2009}{\,} 
\usepackage[version=4]{mhchem}
\usepackage{threeparttable}
\usepackage{array, booktabs}
\title{High voltage carbon-based cathodes for non-aqueous aluminium-ion batteries}
\author{
  Shalini Divya\\
  School of Chemical and Physical Sciences\\
  Victoria University of Wellington\\
  Wellington 6140, New Zealand\\
  E-mail: \texttt{shalini.divya@vuw.ac.nz}\\
   \And
  Thomas Nann\thanks{Corresponding author.}\\
  School of Mathematical and Chemical Sciences\\
  The University of Newcastle\\
  Newcastle, NSW 2308, Australia\\
  E-mail: \texttt{thomas.nann@newcastle.edu.au}\\
}

\begin{document}
\maketitle
\begin{abstract}
 Four different forms of carbon: activated carbon (AC) from human hair, AC from hemp fibers, a carbon fullerene extract consisting of \ce{C60} and \ce{C70} fullerene (CFEx) and Super-P carbon black were tested and compared as cathodes for non-aqueous aluminium-ion batteries (AIBs). These materials differ in their general structure, porosity and morphology. Fullerenes display a crystalline structure, whereas hemp fibers, Super-P and hair are amorphous in nature. Of all materials, AC obtained from hair recorded the highest specific capacity after 50 cycles at 103 mAh g$^{-1}$ with a Coulombic efficiency of $\sim$90\% at a current rate of 50 mA g$^{-1}$. Both hemp fibers and Super-P achieved their highest specific capacities at 56 mAh g$^{-1}$ and 84 mAh g$^{-1}$ respectively. CFEx recorded its highest capacity at 78 mAh g$^{-1}$ and maintained it for 50 cycles. The cells were charged and discharged to 2.45 V and 0.2 V respectively. 
 \end{abstract}
 

\section{Introduction}
Non-aqueous aluminium-ion batteries (AIBs) use low-cost, abundant materials, a non-flammable electrolyte and provide a higher theoretical energy density than the prevalent lithium-ion batteries (LIBs) \cite{lin_ultrafast_2015,paranthaman_transformational_2010,wang_advanced_2017, ambroz_trends_2017}. The aim of this study was to explore systematically how different carbon materials perform as AIB cathodes. Graphite, with its layered structure, is a good intercalation cathode material for many types of ions, including aluminium \cite{ji_recent_2011, yoo_large_2008, lian_large_2010}. We studied other carbon-based materials with the aim to elucidate the energy storage mechanism in these materials. For example, owing to its porous structure, activated carbon (AC) provides a high surface area, and is a high-performing electrode material used in supercapacitors and sodium-ion batteries \cite{eliad_ion_2001, zhu_carbon-based_2011}. In addition to carbonised natural products, we studied fullerenes and a commercial carbon battery additive (Super-P).

\begin{figure}[] 
  \centering
  \includegraphics[width=\textwidth]{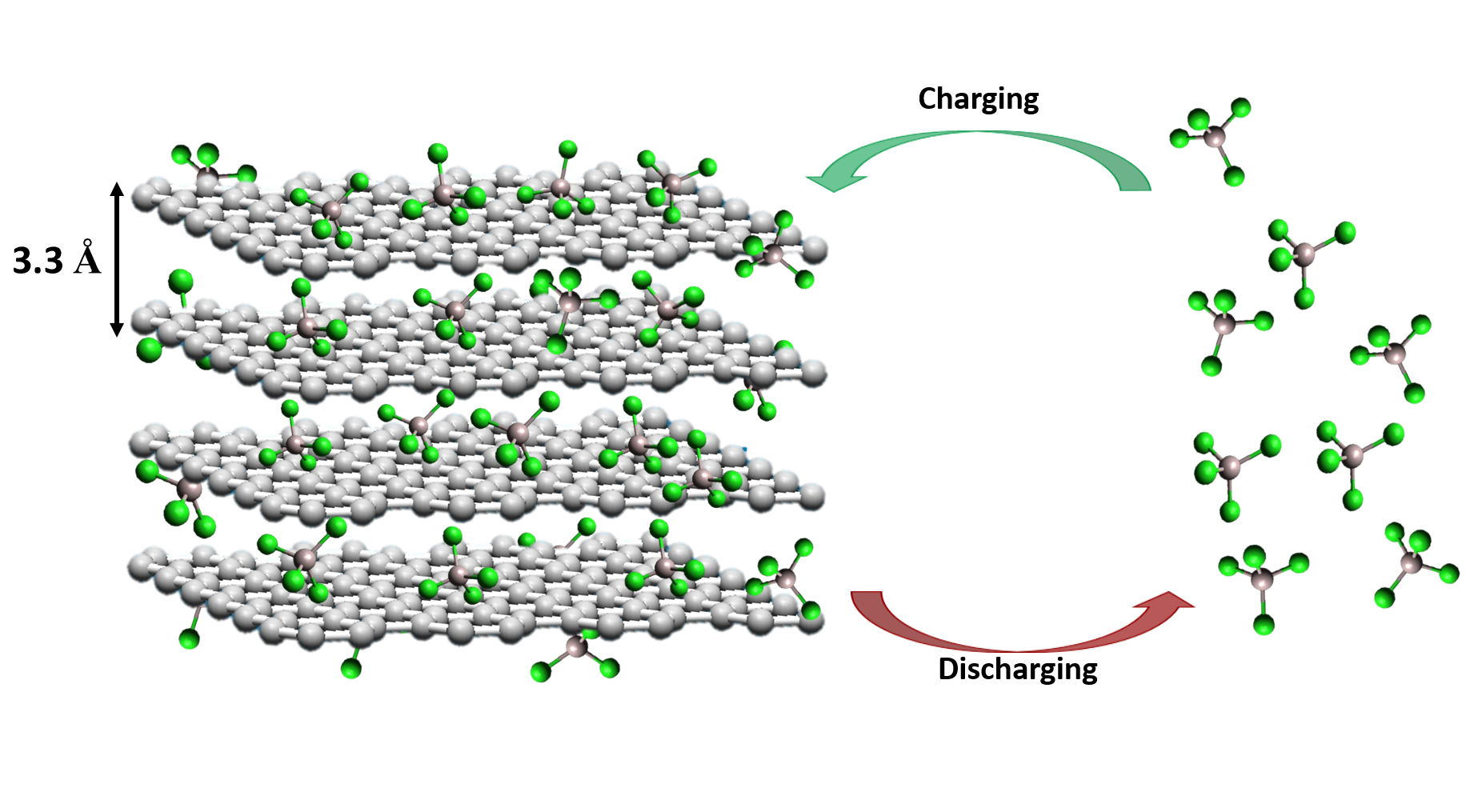}
    \caption{Intercalation and deintercalation of \ce{AlCl4-} ions during charging and discharging in an Al/ graphite cell. The interlayer distance between two graphite sheets is 3.3 \AA.}
  \label{fig:graphitemech}
\end{figure}

Graphite has been repeatedly used as a cathode in AIBs \cite{lin_ultrafast_2015, wang_advanced_2017,zhang_novel_2016}. It displays a high electrical potential {\it vs.} \ce{Al}/ \ce{Al^3+} of 2.1 V. Various forms of graphite such as fluorinated graphite \cite{rani_fluorinated_2013}, kish graphite flakes \cite{wang_kish_2017}, three-dimensional (3D) graphitic-foam \cite{wu_3d_2016}, graphene aerogels \cite{huang_graphene_2019} and many others, have been tested as cathodes for non-aqueous AIBs, showing specific capacities ranging from 60--250 mAh g$^{-1}$. It has been previously established in the literature that \ce{AlCl4-}-anions intercalate into graphitic layers during charge, and deintercalate during discharge as shown in Figure \ref{fig:graphitemech}. X-ray diffraction (XRD) and Raman spectroscopy studies have been used to prove the above \cite{wang_advanced_2017, rani_fluorinated_2013}. Wang \textit{et al.} used Raman spectroscopy to indicate two stages of intercalation in graphite, occurring at two different voltages \cite{wang_advanced_2017}. X-ray photoelectron spectroscopy (XPS) studies also helped in confirming the redox behaviour of carbon during cycling \cite{stadie_zeolite-templated_2017, liu_binder-free_2019,wei_amorphous_2017}, where oxidation occurred during charging and vice-versa. In sodium-ion batteries, graphite has been commonly replaced with hard carbon, \cite{yu_mixed_nodate} since a disordered, and amorphous carbon structure is required for intercalation of bigger sodium ions. Electrodes used in supercapacitors are composed of ACs, typically derived from organic precursors such as rice husk, coconut fibers, etc, due to their moderate cost and high surface area. Their capacitance varies with the mean pore size of the carbon structure and nature of the electrolyte. For example, a capacitor using a non-aqueous electrolyte displays a higher capacitance \cite{simon_materials_2008}. In addition, removal of moisture during heat treatment enhances the stability of the carbon structure, which subsequently improves its cycle life. Naturally, carbon-based materials are the front runners when selecting an electrode material for any new energy storage device. 
Here, four different carbon-based materials --- AC from hair, AC from hemp fibers, fullerene extract (CFEx) and Super-P were investigated as cathodes for AIBs. AC derived from both hair and hemp fibers contain pores of various sizes (mesopores and micropores). Fullerenes, formed entirely by the covalently bonded carbon atoms, displayed a tightly packed structure with an approximate Brunauer-Emmet and Teller (BET) surface area of 19 m$^2$ g$^{-1}$ \cite{yuan_coupling_2016}. Fullerenes have been tested as electrodes in various energy storage devices \cite{chabre_electrochemical_1992, sood_electrochemical_2018, zhang_fullerenes_2014}, but never before as a cathode for AIBs. Super-P is an amorphous form of carbon, which is mainly used as an additive to enhance the conductivity of an electrode slurry. With a highly disordered structure \cite{see_reversible_2017}, and a surface area of 62 m$^2$ g$^{-1}$, its pore size ranges from $\sim$30 to 50 nm \cite{younesi_analysis_2015}. To determine if the carbon additive contributed any capacity of its own, Super-P was tested as an active cathode material in this work. 
\section{Results and discussion}
\begin{figure}
  \centering
  \includegraphics[width=\textwidth]{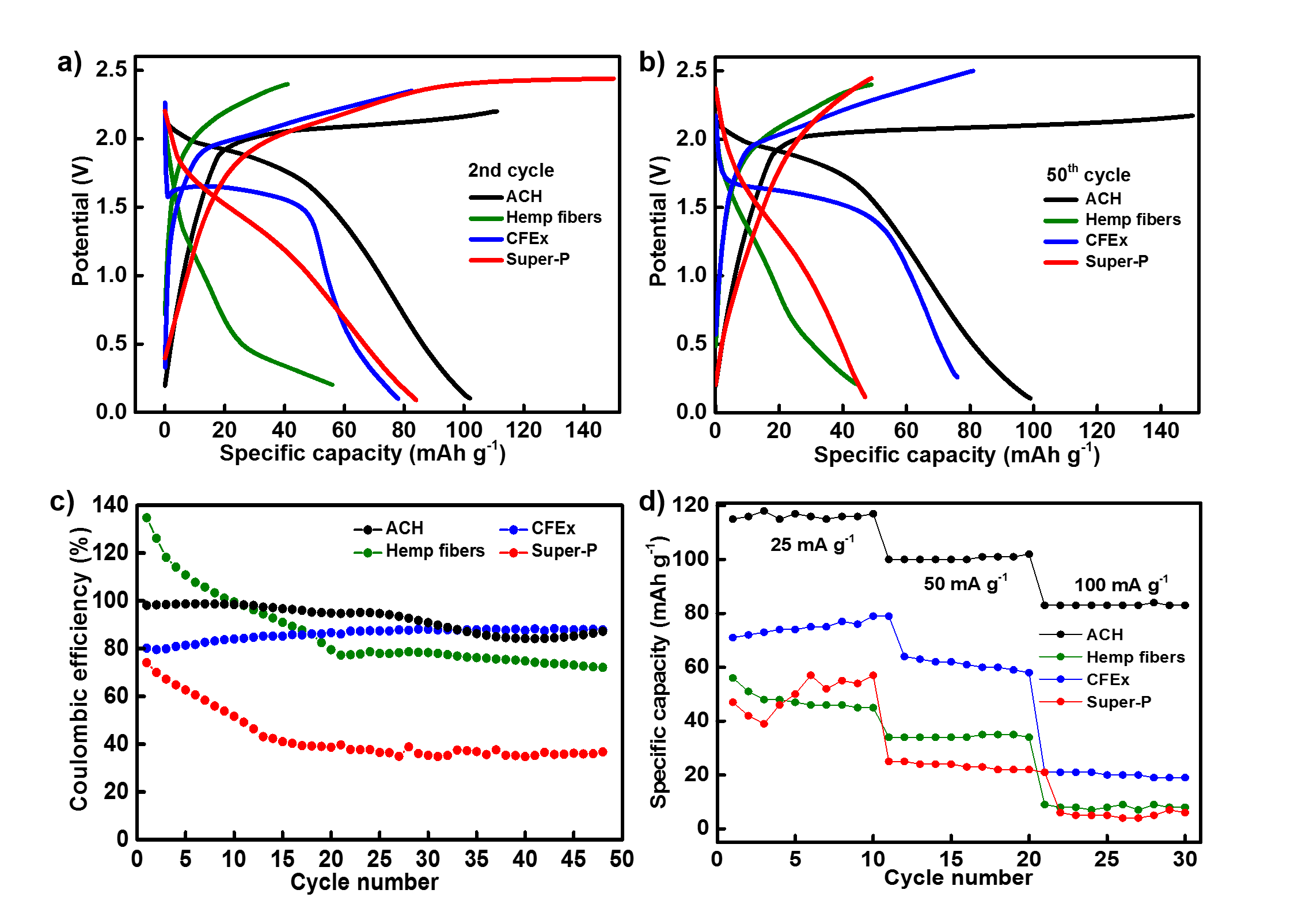}
    \caption{Specific capacities of AC (human hair, hemp fibers), CFEx and Super-P in their a) first and b) 50$^{th}$ cycle at a current rate of 50 mA g$^{-1}$. c) Coulombic efficiencies (CEs) of cells at a current rate of 50 mA g$^{-1}$. d) Galvanostatic charge/ discharge profile of all cells at various current rates ranging from 25 mA g$^{-1}$ to 100 mA g$^{-1}$ in a two-electrode setup against Al.}
  \label{fig:CDCall}
\end{figure}

The battery system used in this study comprised a cathode, 99\% pure Al foil as the anode, and a room temperature ionic liquid (RTIL) as electrolyte. Using a battery analyser, specific capacities and Coulombic efficiencies (CEs) of the cathodes were recorded at various current rates (cf.\ Figure \ref{fig:CDCall}). Cathode morphologies before and after the cycles were studied using Raman spectroscopy, X-ray diffraction (XRD) patterns and scanning electron microscopy (SEM).

\begin{figure}
  \centering
  \includegraphics[width=\textwidth]{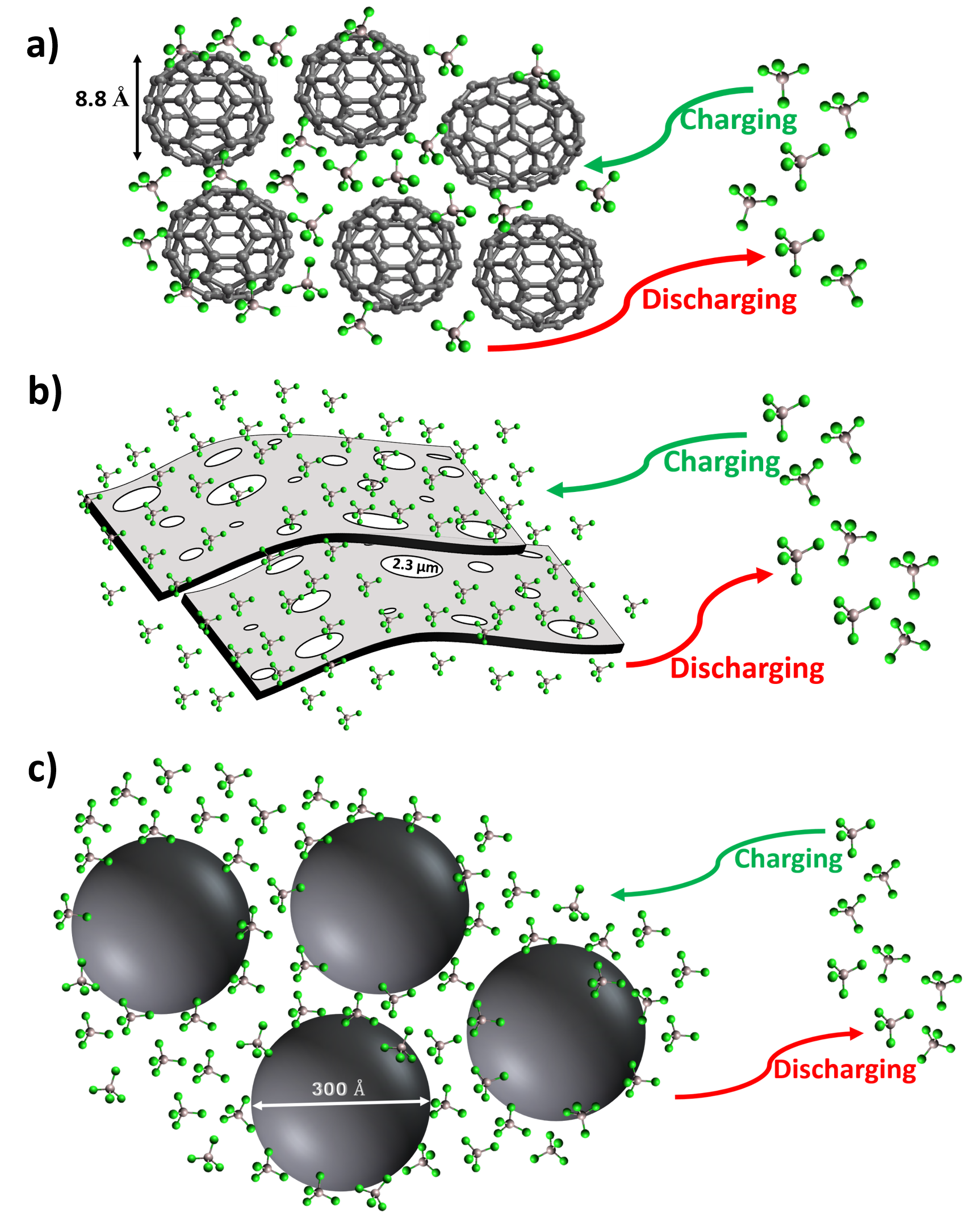}
    \caption{Suggested mechanism for an a) Al/ CFEx cell, b) Al/hemp cell, hemp fibers have pore sizes as large as 2.0--2.5 $\mu$m allowing the \ce{AlCl4-} to get absorbed onto their surface. However, agglomeration of these fibers after a few cycles reduced the number of active sites available for effective charge storage, and c) Al/Super-P cell, chloroaluminates intercalate into the very few graphitic planes in Super-P, while few anions adsorb onto its surface. Further cycling leads to cathode pulverisation, which results in capacity fading.}
  \label{fig:allmech}
\end{figure}

\begin{figure}
  \centering
  \includegraphics[width=\textwidth]{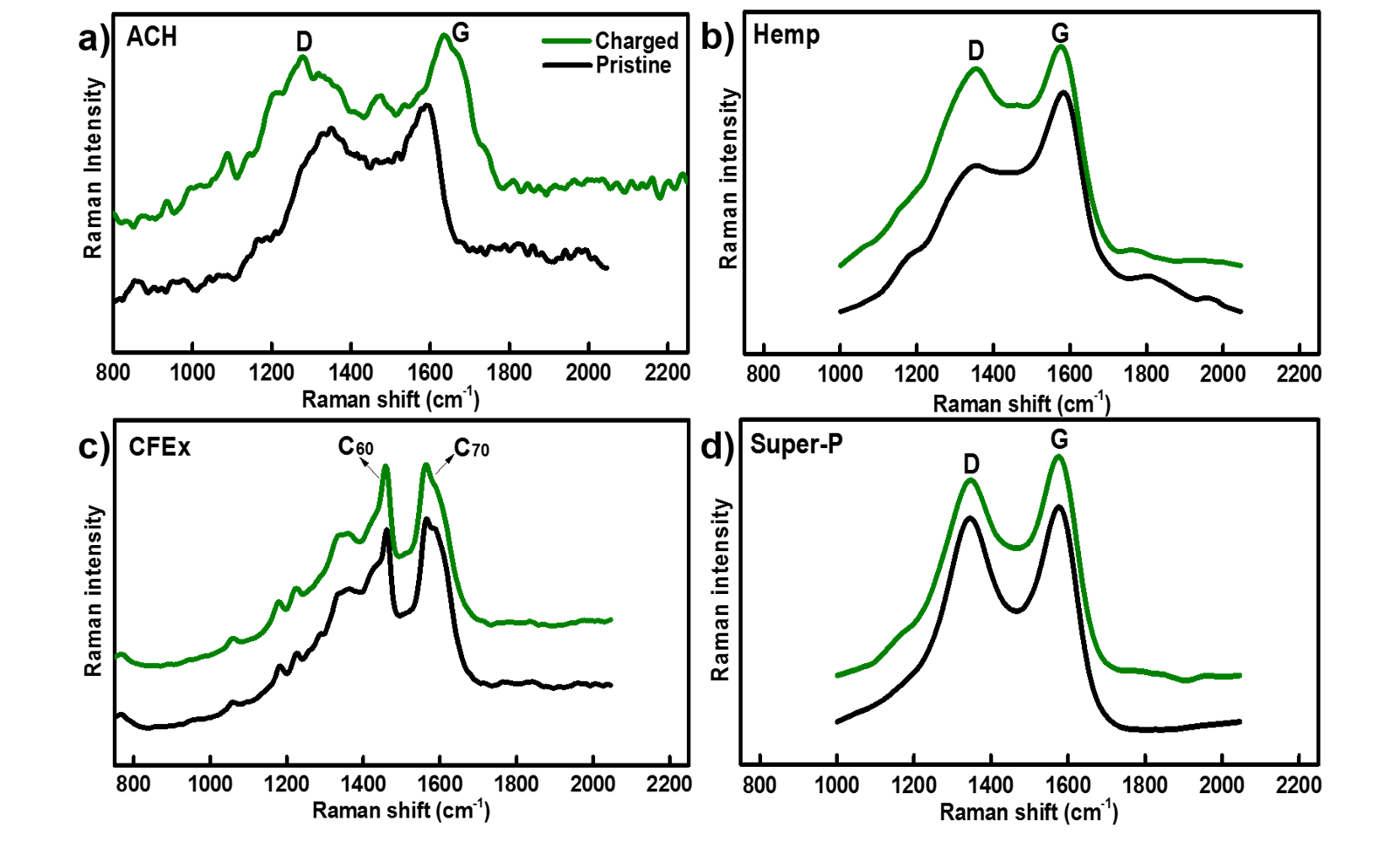}
    \caption{Raman spectra of pristine (black) and charged (green) a) AC from hair (ACH), b) hemp fibers, c) CFEx and d) Super-P cathodes showing the presence of both D and G bands.}
  \label{fig:raman}
\end{figure}

Figures \ref{fig:CDCall}a and \ref{fig:CDCall}b show the specific capacities of all cells for their first and 50$^{th}$ cycles. CE of cells at a current rate of 50 mA g$^{-1}$, and cycling performances under different current rates ranging from 25 to 100 mA g$^{-1}$ are displayed in Figure \ref{fig:CDCall}c and \ref{fig:CDCall}d. Hair cathodes, in black, exhibited a high capacity of $\sim$100 mAh g$^{-1}$ with CE of $\sim$95\%. Hemp cells, in green, displayed lower capacities starting from 56 mAh g$^{-1}$ in their first cycle, which decreased to 45 mAh g$^{-1}$ after 50 cycles. CFEx, in blue, showed a capacity of $\sim$80 mAh g$^{-1}$, which was maintained for 50 cycles, with an average CE of 90\% . With an initial value of 84 mAh g$^{-1}$, the specific capacity of Super-P, shown in red, decreased to 47 mAh g$^{-1}$ and a low CE of $\sim$40\% was observed. It was noted that the CEs and the specific capacities of hemp fibers and Super-P cells decreased considerably after repeated cycles. 
It was interesting to observe that hair and hemp seemed to follow a different mechanism, although they were both activated forms of carbon. While hair maintained a capacity of over 100 mA g$^{-1}$ over 50 cycles, hemp cells failed to retain 50\% of their initial capacity during the cycles. CE of the hemp fibers also showed a decreasing trend (CE < 40\%) unlike hair (CE $\sim$ 90\%). To examine this behavior, Raman spectroscopy was used and cycled (charged) cathodes of all materials were compared with the pristine ones. The results are illustrated in Figure \ref{fig:raman}. G-band, present at 1600 cm$^{-1}$ in carbon structures, commonly indicates a graphitic structure. In addition, a D-band, observed at 1300 cm $^{-1}$ indicates presence of some structural disorder. The existence of G-band in hair (Figure \ref{fig:raman}a), hemp (Figure \ref{fig:raman}b) and Super-P (Figure \ref{fig:raman}d) revealed the presence of graphitic planes. D-bands were expected in the spectra since the materials were majorly composed of amorphous carbon. Comparing hair and hemp, it was noted that the D-band's intensity in hemp increased significantly after cycles, suggesting dislocation of a few graphene sheets resulting in structural disorder, shown in Figure \ref{fig:raman}b. 

\begin{figure}
  \centering
  \includegraphics[width=\textwidth]{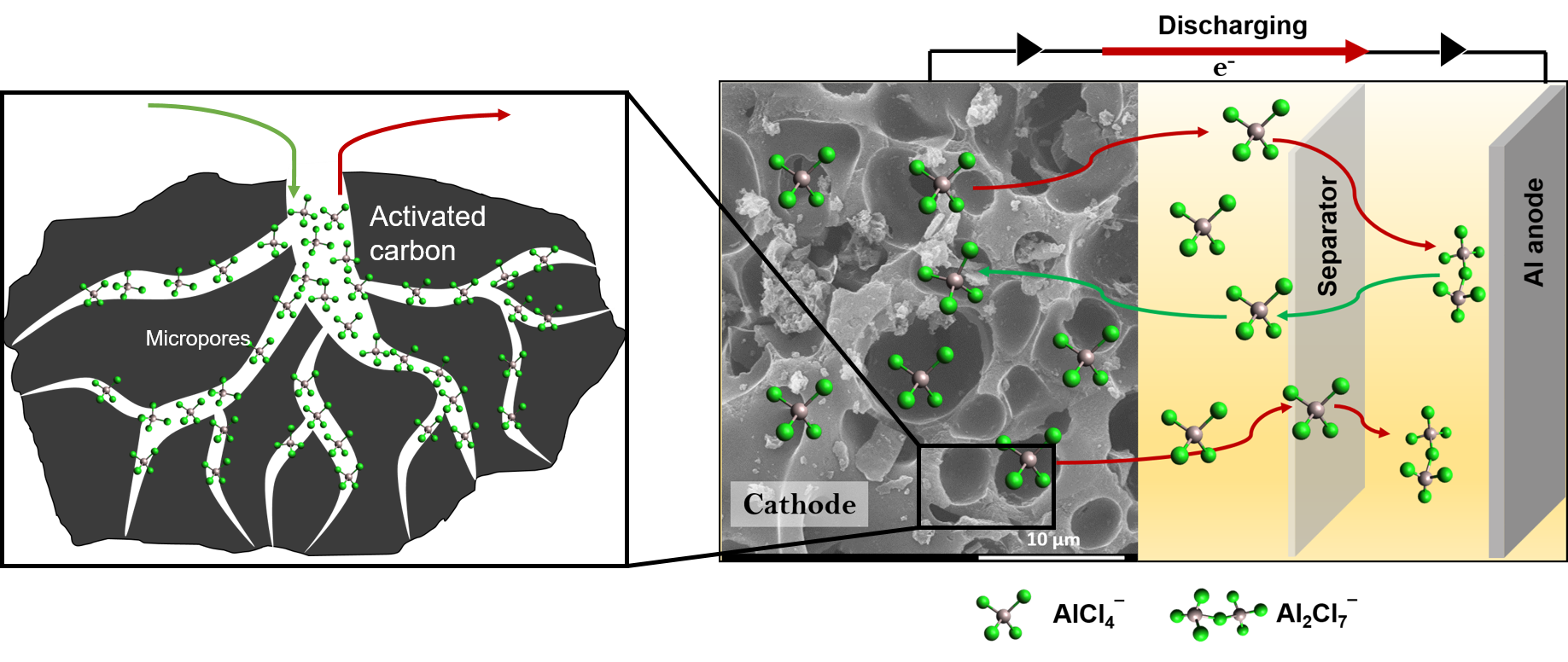}
    \caption{Suggested mechanism for an Al/hair cell. Chloroaluminate ions (\ce{AlCl4-}) intercalate into the few graphitic planes and micro/ mesopores present in them, in addition to surface adsorption of ions displaying both Faradaic and non-Faradaic processes for charge storage.}
  \label{fig:ACHmech}
\end{figure}

In Raman spectroscopy, an increased full-width at half maxima (FWHM) suggests an increase in the lattice defects and structural deformities. In Figure \ref{fig:raman}, the presence of these defects may validate intercalation of chloroaluminate ions in the few graphitic layers present in the carbon matrix of hair, hemp fibers and Super-P cathodes. Further, a highly porous structure that allows surface adsorption of charge carriers, much like in supercapacitors \cite{beguin_carbons_2014}, might also result in an uneven lattice structure. Since AC lacks a long-range order, cell capacity derived exclusively from intercalation can be ruled out. It can be assumed that an additional surface-based adsorption of chloroaluminates contributed to the total cell capacity \cite{brezesinski_ordered_2010}. The superior performance of Al/hair batteries can be attributed to this dual charge-storage mechanism, illustrated in Figure \ref{fig:ACHmech}.

Raman spectra of CFEx in Figure \ref{fig:raman}c show the characteristic bands of both \ce{C60} and \ce{C70} molecules. A `pentagonal pinch mode', usually observed in a \ce{C60} spectrum, was present at 1460 cm$^{-1}$. \ce{C70} molecules resulted in multiple bands due to reduced molecular symmetry, which increased the number of vibrational modes, consequently increasing the number of active Raman bands \cite{kimbrell_analysis_2014}. The spectra of charged CFEx was strikingly similar to the pristine ones. Since Raman spectroscopy is sensitive to minute differences in the molecular morphology, the results suggested that the fullerenes did not undergo any substantial changes during the cycles. 

\begin{figure}
  \centering
  \includegraphics[width=0.8\textwidth]{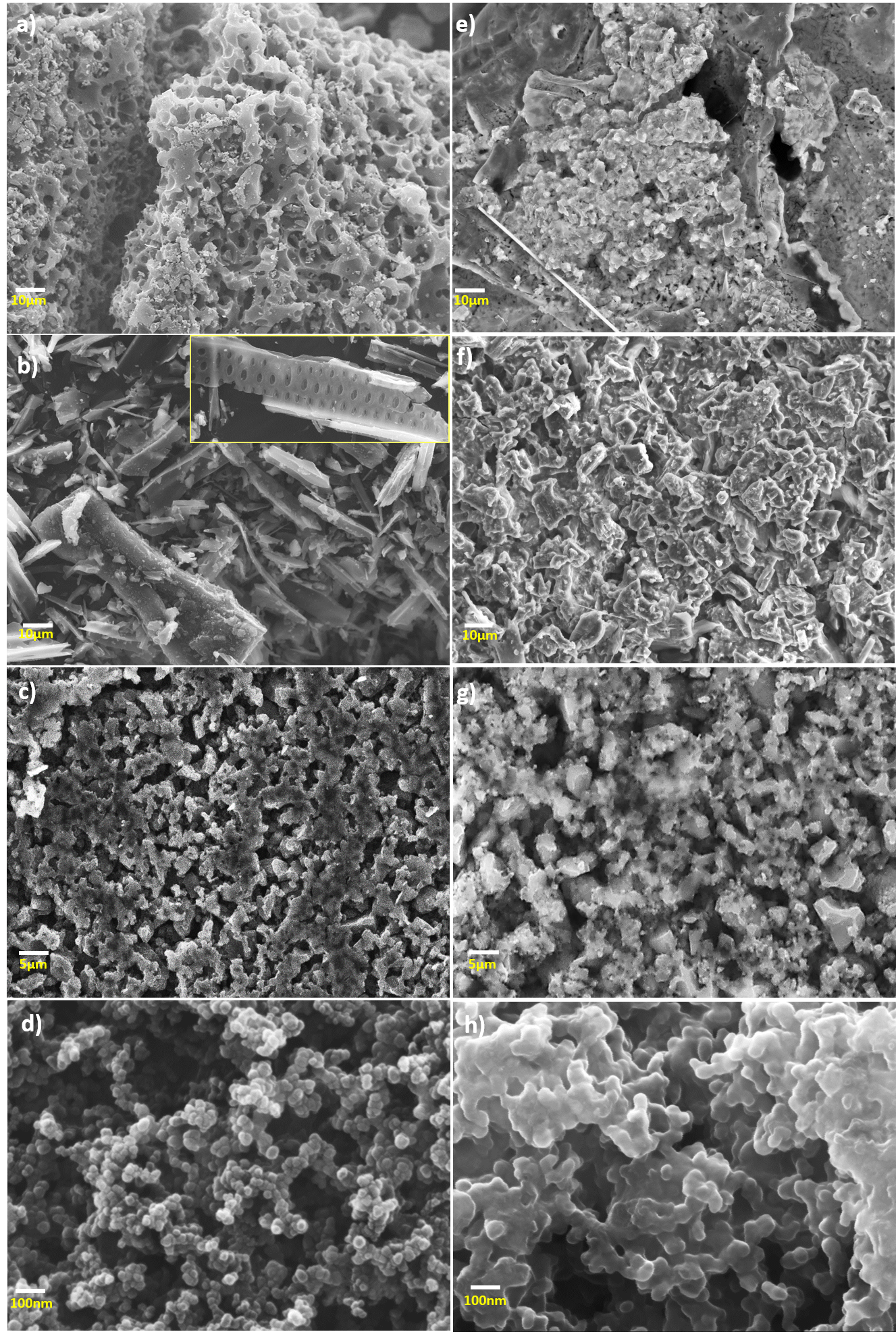}
    \caption{Scanning electron microscopy (SEM) images comparing pristine a) human hair, b) hemp, c) CFEx and d) Super-P and charged e) human hair, f) hemp, g) CFEx and h) Super-P cathodes. Hemp fibers and Super-P undergo permanent changes after charge/ discharge cycles and fail to retain capacity.}
  \label{fig:SEM}
\end{figure}

To further understand the durability of the CFEx batteries, and the unstable nature of hemp and Super-P, electron microscopy was used. Figure \ref{fig:SEM} shows the SEM images of pristine (Figure \ref{fig:SEM}a, b, c and d) and charged (Figure \ref{fig:SEM}e ,f ,g and h) cathodes of all carbon materials. Although, fullerene retained its surface morphology after 30 cycles, it was observed that hemp and Super-P underwent significant changes. Hemp lost its surface porosity, as shown in Figure \ref{fig:SEM}f, while Figure \ref{fig:SEM}d and h implied that Super-P agglomerated, and presumably sintered together. Recently, Canever \textit{et al.} reported that the presence of surface defects on a cathode plays a major role in the formation of the solid electrolyte interface (SEI), which ultimately leads to a cathode's poor capacity retention. \cite{canever_solid-electrolyte_2020}. It can be concluded that the absence of a long-range order in hemp and Super-P might be a contributing factor to the formation of an SEI layer, resulting in their poor performance as a battery material. Moreover, both hemp fibers and Super-P have a highly disordered structure to begin with. Repeated intercalation or absorption of ions on their surface caused further damage, and the batteries failed to retain their capacity and displayed low CEs (cf.\ Figure \ref{fig:CDCall}a and b). The reduced capacity and low CE may also be a result of certain side reactions. These may include interaction of the electrode or electrolyte with certain impurities or degradation of the cathode structure (pulverisation) \cite{gyenes_understanding_2015}. 

Figure \ref{fig:SEM}c and g show the topography of a CFEx cathode, before and after cycling. The micrographs do not contain any significant information. It is known that fullerenes have a cage-like morphology \cite{kroto_c60_1985}. However, the chloroaluminates are bigger in size than the cage cavities, and are unable to squeeze in and out of them. Nonetheless, charge-storage on the outer surface of fullerenes is always a possibility. It is hypothesised that \ce{AlCl4-} anions migrated through the gaps present in between two fullerenes and the surface-based redox processes took place in a systematic way throughout the cathode surface. This way, the fullerene structure remained intact with no significant lattice disorder after the electrochemical processes; which is precisely what was observed in the Raman spectra and SEM images. Consequently, fullerenes were able to store charge reversibly. This finding points at a mechanism different from ion intercalation. Since CFEx lacks graphitic planes to intercalate ions, surface adsorption of ions is highly likely \cite{adams_van_1994}. A schematic of the mechanism is shown in Figure \ref{fig:allmech}a. 

\begin{figure}
  \centering
  \includegraphics[width=\textwidth]{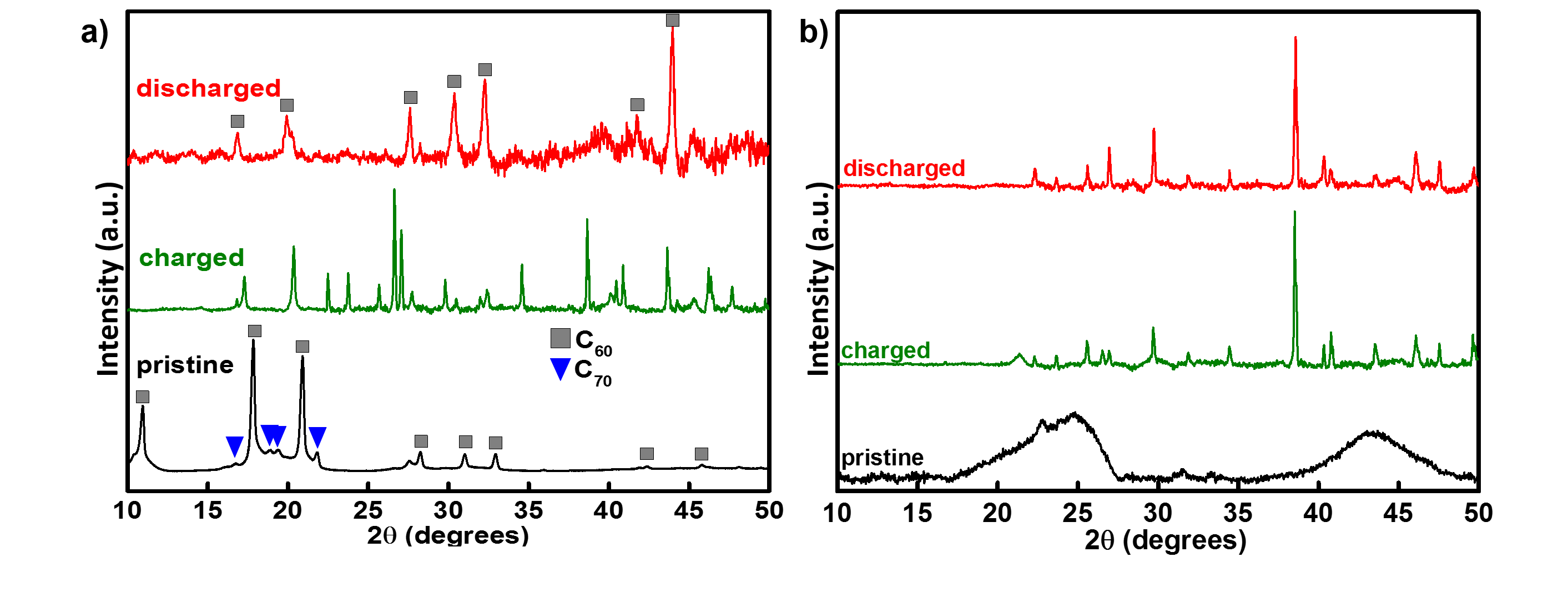}
    \caption{X-ray diffraction (XRD) patterns of a) CFEx and b) human hair cathodes to study changes in their lattice after galvanostatic cycles in a two-electrode setup against Al with characteristic peaks marked for \ce{C60} (grey box) and \ce{C70} (blue inverted triangle).}
  \label{fig:XRD}
\end{figure}

\begin{figure}
  \centering
  \includegraphics[width=\textwidth]{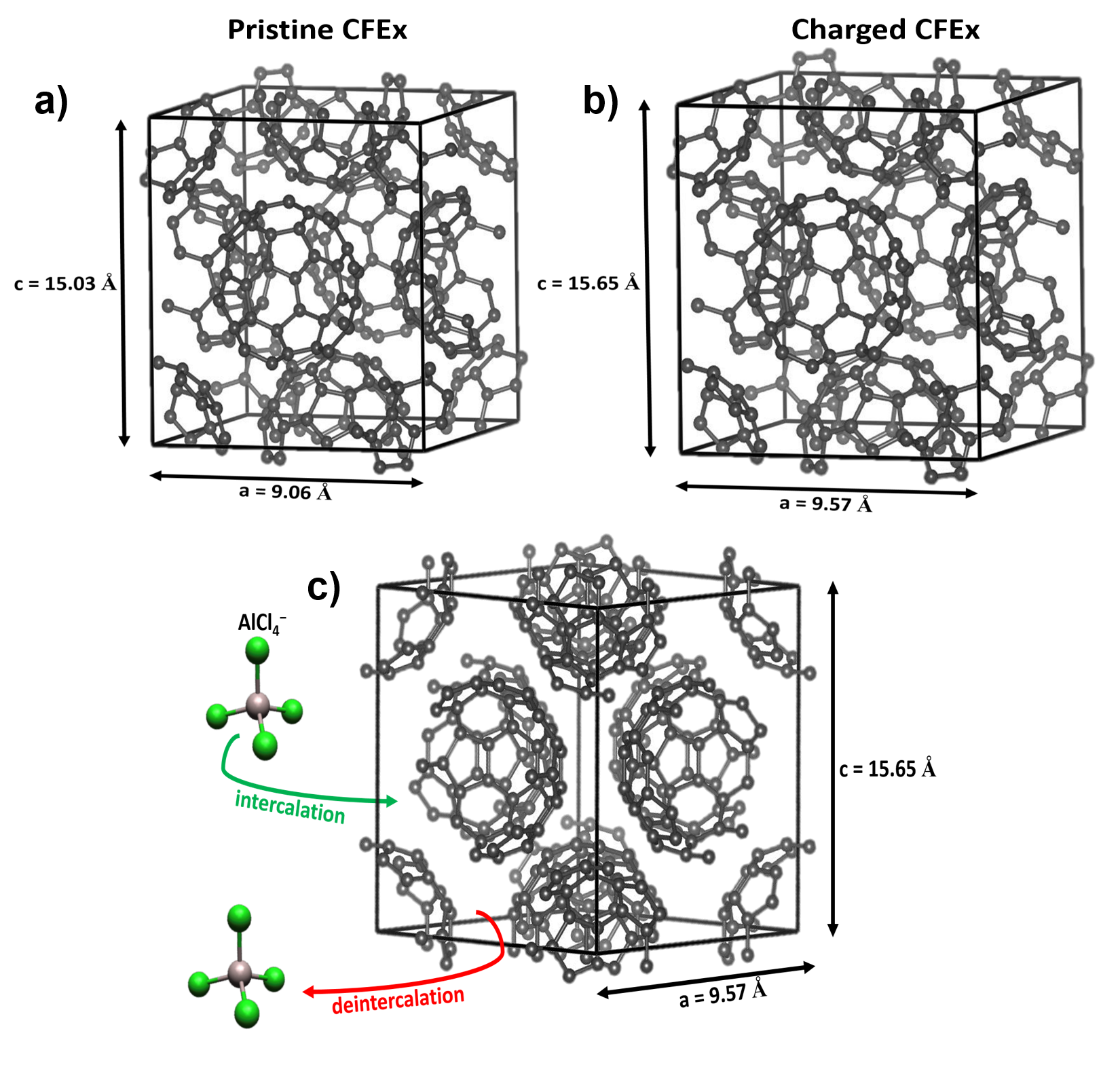}
    \caption{Changes in the lattice parameters of a \ce{C60} unit cell. a) Pristine \ce{C60} unit cell, b) charged \ce{C60} unit cell with increased parameters suggesting a uniform shift in the lattice after charge/discharge. c) Expected intercalation sites of \ce{AlCl4-} ions in the unit cell.}
  \label{fig:cfexcrys}
\end{figure}

X-ray diffraction patterns were studied to elucidate the mechanism of hair and CFEx cells. Pristine (black), charged (green) and discharged (red) cathodes, shown in Figure \ref{fig:XRD}a and b, represent hair and CFEx respectively after 30 charge/discharge cycles. Figure \ref{fig:XRD}a displays the characteristic peaks of both \ce{C60} and \ce{C70} at 2$\theta$ values of 10.9$^{\circ}$, 17.8$^{\circ}$, 20.9$^{\circ}$ and 28.2$^{\circ}$ for \ce{C60}, and 18.9$^{\circ}$, 19.3$^{\circ}$ and 21.8$^{\circ}$ for \ce{C70} molecules respectively. In addition to the new diffraction peaks that appeared at lower 2$\theta$ values, a few peaks shifted from their original 2$\theta$ values in the charged CFEx cathode. However, after discharge, the XRD peaks returned to their original peak positions. This phenomenon strongly suggested a reversible process taking place in CFEx. To confirm this, the unit cell lattice parameters for both pristine and charged cathodes for a \ce{C60} molecule were calculated. The unit cell had a tetragonal crystal system with space group of \textit{P42/mmc} and space group number 131 (ICDD: 04-013-1339). Lattice parameters `a' and `b' for the charged cathode increased from 9.06 \AA\ to 9.57 \AA\ and `c' increased from 15.03 \AA\ to 15.65 \AA, as shown in Figure \ref{fig:cfexcrys}a and b. Lattice parameters of the discharged fullerene cathode were closer to the pristine values. These changes suggested a reversible mechanism, where chloroaluminates entered the free spaces between two fullerene molecules. A possible site for \ce{AlCl4-} intercalation is depicted in Figure \ref{fig:cfexcrys}c. 

Unfortunately, XRD patterns of hair cathodes were inconclusive (Figure \ref{fig:XRD}b). Pristine cathodes show broad peaks that confirmed a structure that was amorphous and highly porous. Yet, the material became more symmetrical and crystalline after cycling. Charged and discharged cathodes exhibited similar looking patterns implying no change in the newly formed crystal lattice. It has to be noted that the presence of crystallinity in an active material does not limit the surface-based charge storing capacity \cite{kim_synthesis_2006, jow_factors_2018}. Therefore, possibility of a reversible intercalation of chloroaluminates into the graphitic planes cannot be ruled out. Further analysis is required to investigate this unique dual behaviour and establish the mechanism of the Al/hair battery.

\begin{figure}
  \centering
  \includegraphics[width=0.85\textwidth]{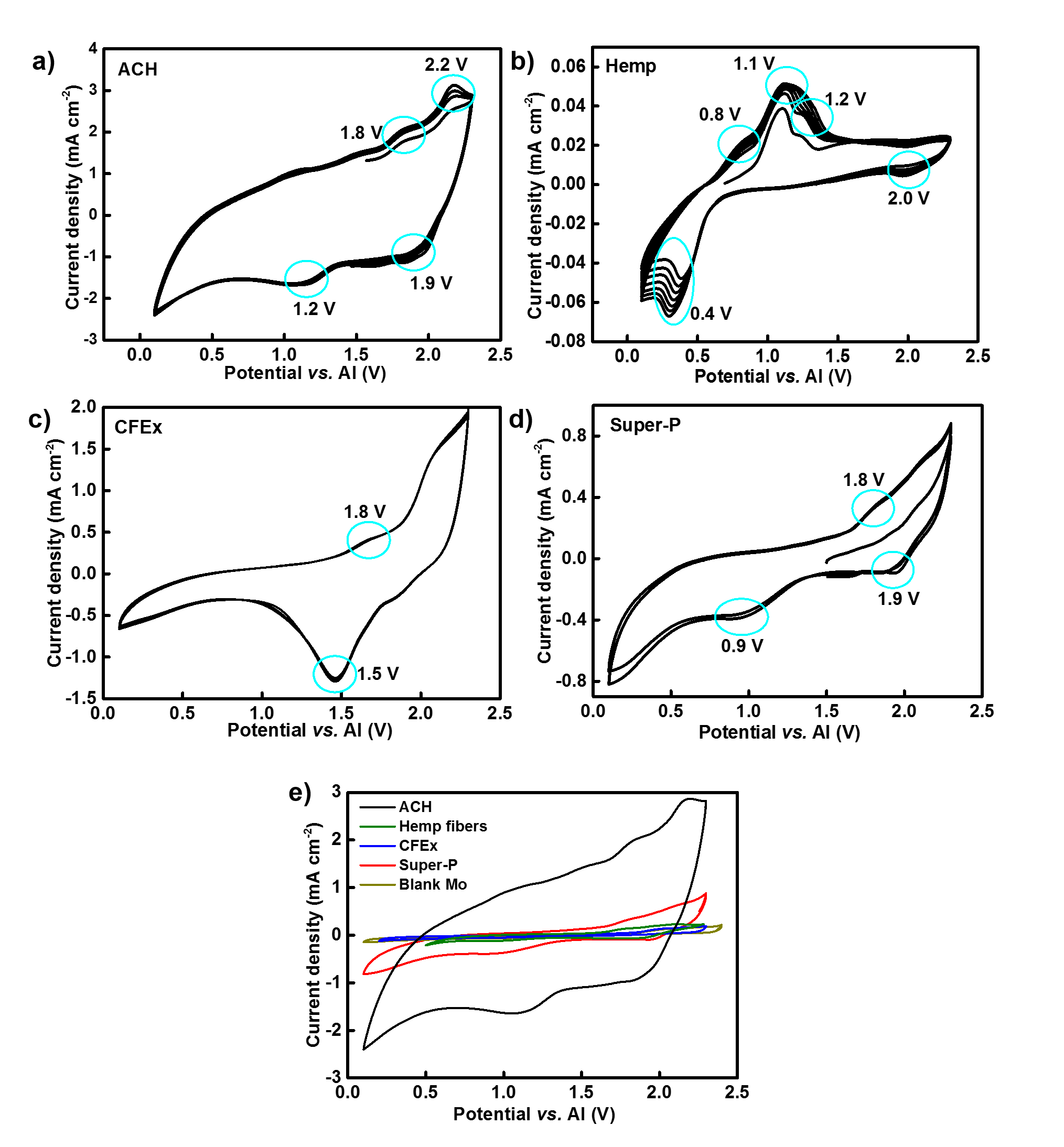}
    \caption{Cyclic voltammograms of a) AC from hair (ACH), b) hemp fibers, c) CFEx and d) Super-P cathodes at a scan rate of 10 mV s$^{-1}$ against Al as the counter/reference electrode in a two-electrode setup. ACH cathode observed a larger CV area than other cathodes, which comes from an additional pseudocapacitance, adding capacity to the system.}
  \label{fig:CV}
\end{figure}

\begin{figure}
\centering
\includegraphics[width=0.8\textwidth]{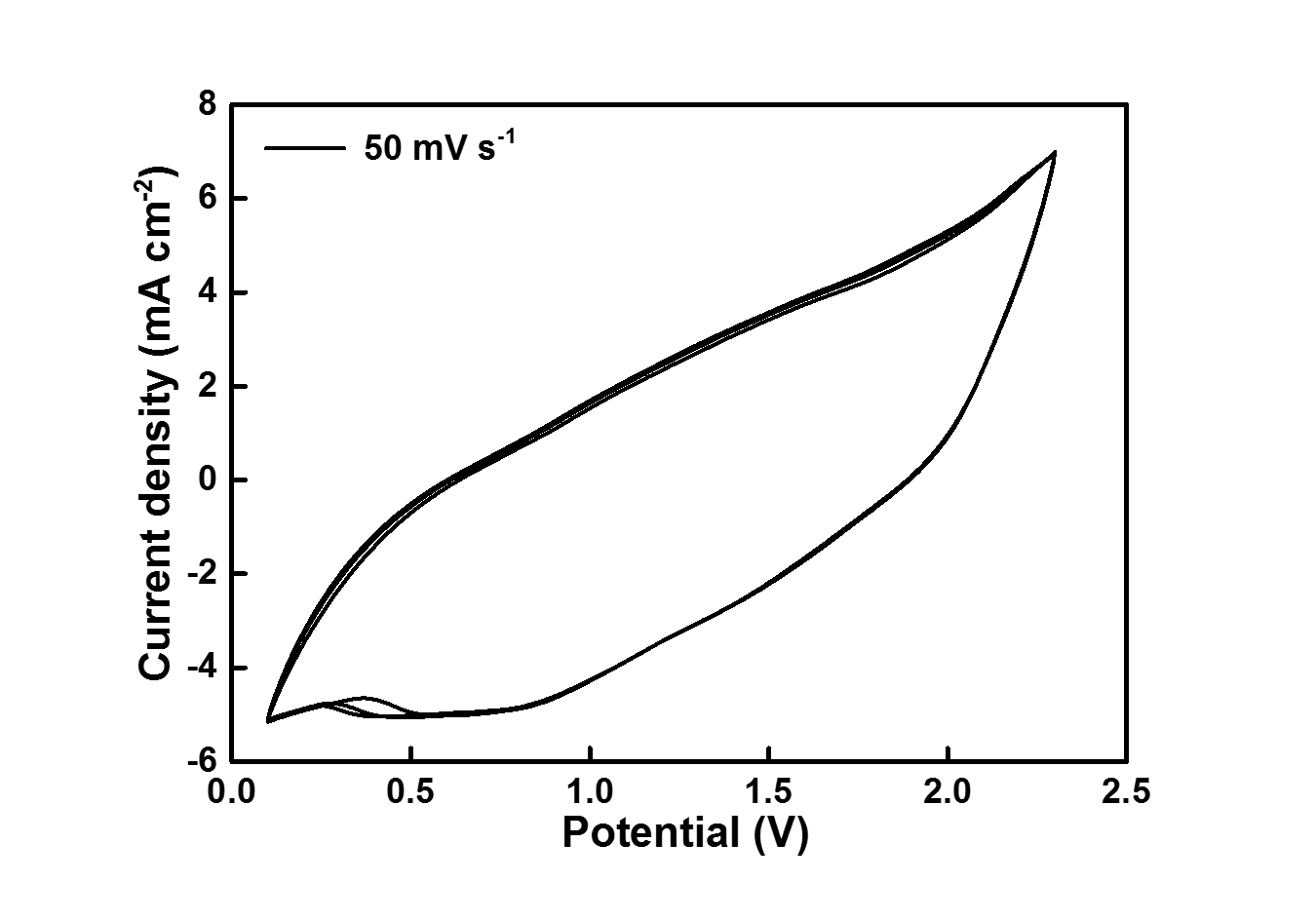}
\caption{Cyclic voltammogram of ACH at a scan rate of 50 mV s$^{-1}$ in a two electrode setup against Al showing a capacitor-like behaviour with no visible oxidation-reduction peaks unlike Figure \ref{fig:CV}c, where redox peaks were observed.}
\label{fig:hair50mVs}
\end{figure}

\begin{figure}
  \centering
  \includegraphics[width=\textwidth]{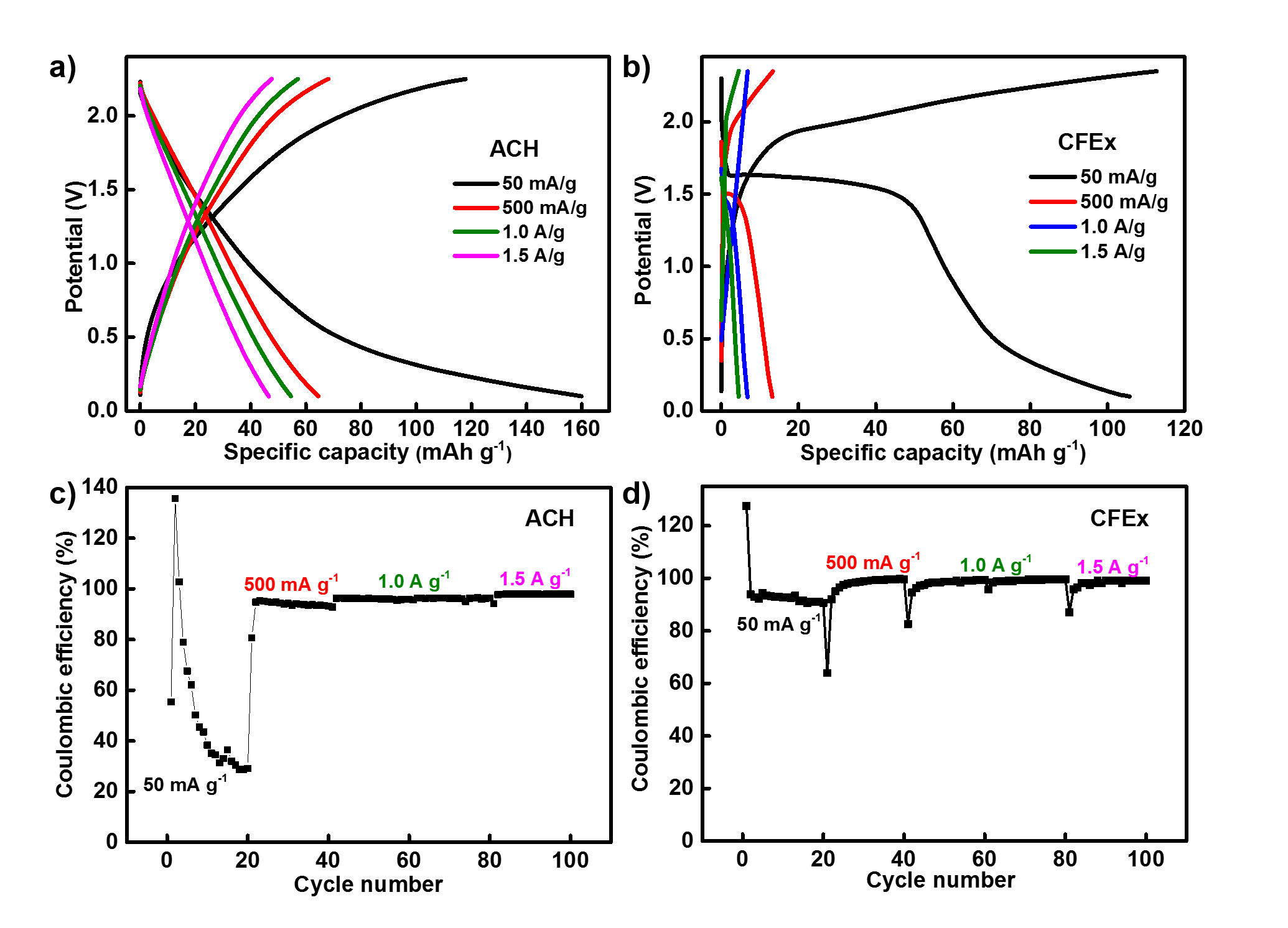}
    \caption{Discharge capacities of a) ACH and b) CFEx cathodes at current rates of 50 mA g$^{-1}$, 500 mA g$^{-1}$, 1.0 A g$^{-1}$ and 1.5 A g$^{-1}$ along with their CEs.}
  \label{fig:CFExACHlong}
\end{figure}

Lastly, to verify whether hair was acting as a pseudocapacitive material that allowed dual charge-storage mechanism, cyclic voltammograms of all cathodes at a scan rate of 10 mV s$^{-1}$ were compared. Figure \ref{fig:CV}a--e showed that the hair batteries resulted in a more rectangular, capacitor-like cyclic voltammetry (CV) curve. However, redox processes were noticeable at a lower scan rate of 10 mV s$^{-1}$ and small redox peaks were visible in Figure \ref{fig:CV}b at 1.8 and 2.1 V during charging, and 1.1 and 1.9 V during discharge, matching well with the charge and discharge plateaus observed in Figure \ref{fig:CDCall}a and b. At a higher scan rate of 50 mV s$^{-1}$, the redox peaks disappeared and the material displayed an ideal capacitor-like CV curve, shown in Figure \ref{fig:hair50mVs}, displayed in the SI. This happens because at higher scan rates, the diffusion layer gets thinner and more ions are in close proximity of the electrode \cite{guan_capacitive_2016, dupont_separating_2015}. As a result, higher currents were recorded and the intensity of the redox peaks became negligible. Although redox peaks were observed for CFEx (at 1.8 V during charging and 1.5 V during discharging), corresponding to the discharge plateau at $\sim$1.5 V in Figure \ref{fig:CDCall}a and b and others (Super-P: 1.8 V during charge, and 1.9 and 0.9 V during discharge, hemp fibers: 1.1, 1.2 and 0.8 V during charge, and 1.98 and 0.38 V during discharge), the measured currents for the Super-P and hemp fibers were very low (negligible compared to hair). At such low currents, it becomes impossible to draw conclusions about the redox reactions taking place and the stability of the species resulting from the electron transfer. Figure \ref{fig:CFExACHlong} in the SI, shows the 50th cycle measurement for Al/hair and Al/natural graphite cells. The hair batteries not only display a higher specific capacity than conventional graphite, but also a high voltage of 1.92 V with an energy density of 202 Wh kg$^{-1}$.

\clearpage

\begin{table}
\caption{Comparing battery metrics of all carbon-based cathodes tested in this work} \label{table1}
\begin{center}
\begin{tabular}{|lccc|}
\hline
Active material  & {\textbf{Specific capacity}} & {\textbf{Cell efficiency}} & {\textbf{Cell voltage}}\\
 & {\textbf{(mAh g$^{-1}$)}} & {\textbf{(\%)}} & {\textbf{(V)}}\\
\hline
Human hair &  102 & 97 & 1.9 \\
Fullerene mix &  78 & 85 & 1.7 \\
Hemp fibers & 49 & 75 & 1.8 \\
Super-P & 46 & 40 & 1.5 \\
\hline  
\end{tabular}
\end{center}
\end{table}

\section{Conclusions}
In summary, AC derived from hair proved to be the best carbon-based cathode among all the tested materials, with a specific capacity of $\sim$100 mAh g$^{-1}$ at a potential of 1.9 V and a CE of $\sim$90$\%$ together with a dual charge-storage mechanism in place. Hemp fibers and Super-P have a highly amorphous structure, and underwent structural degradation after a few cycles, resulting in low capacity values. The presence of functional groups on their surface further deteriorated its capacity retention. In CFEx, \ce{AlCl4-} anions reversibly migrate in and out of the gaps in between the fullerenes. The XRD data suggests that the crystal slightly expands and contracts  during charge/discharge. Moreover, fullerenes maintain their structural integrity and CE throughout the cycles. The high performance of the Al/hair battery can be attributed to the high porosity of the material combined with the presence of hetero-atoms on its surface, which enhanced its wettability. The cumulative effect of the non-Faradaic and Faradaic processes lead to the material achieving a higher capacity. The battery metrics of all cathodes tested in this work have been listed in Table \ref{table1}.


\section{Experimental section}
\subsection{Chemicals}
\subsubsection*{AC from human hair}
\begin{figure}[ht!]
\centering
\includegraphics[width=\textwidth]{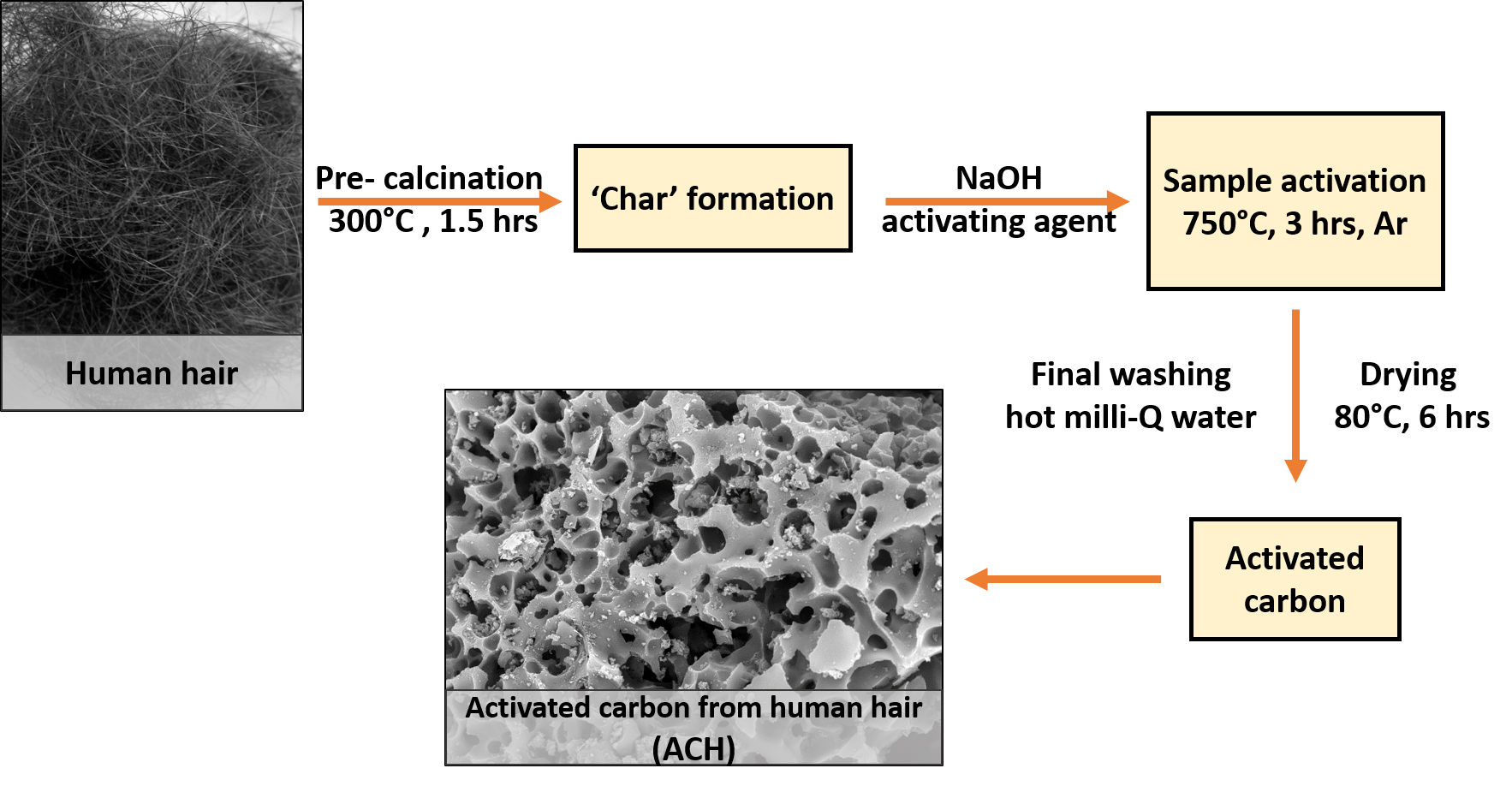}
\caption{Synthesis of AC from human hair using NaOH as the activating agent.}
\label{fig:ACHsyn}
\end{figure}

Human hair was donated by a colleague from the Nann group. No ethics approval was required. The hair was initially cut into small pieces and washed thoroughly using isopropanol (IPA) to get rid of any chemical residue (from shampoo or conditioners). The samples were then dried at 100 $^{\circ}$C for 12 hours. The dried hair was then burnt in Ar atmosphere at 300 $^{\circ}$C for 90 minutes. The pre–carbonized sample was then mixed with NaOH (Sigma-Aldrich) in a mass ratio of 1:2. The sample was later carbonized at 750 $^{\circ}$C for 3 hours in presence of Ar. The product obtained was repeatedly washed with hot de-ionised water to remove any traces of sodium from the sample, which was then dried at 80 $^{\circ}$C for 6 hours to obtain the final product. In this reaction, NaOH was reduced to free metal, Na. These atoms in turn expanded the carbon matrix after intercalating into the carbon structure. Increased temperature (750$^{\circ}$ C) forced the  atoms out of the carbon matrix, thus creating micropores. Oxidation of carbon from oxygen atoms of the hydroxide group formed carbon dioxide (\ce{CO2}), providing routes for channeling the sodium atoms into the internal structure, resulting in a well-connected porous structure \cite{satish_macroporous_2015}. The calcinating temperature used here was 750$^{\circ}$ C. Figure \ref{fig:ACHsyn} displays a flowchart describing the ACH synthesis. 

\subsubsection*{AC from hemp fibers}
The material was provided by Carbon Valley and used as received.

\subsubsection*{Fullerene extract}
\ce{C60}/ \ce{C70}, approx. 85\% \ce{C60}, 14\% \ce{C70}, and 1\% higher fullerenes, was purchased from SES research and used as received.

\subsubsection*{Super-P carbon black}
Super-P conductive carbon, 99+\% metals basis was purchased from Alfa Aesar and used as received

\subsection{Cathode preparation}
A slurry was prepared by mixing the active material (85$\%$ by wt.), 9$\%$ binder (PVDF, MTI Corp.) and 6$\%$ Super-P conductive carbon (99+$\%$ metals basis, Alfa Aesar) in N-methyl pyrrolidone NMP (anhydrous, 99.5$\%$, Sigma-Aldrich). To form the Super-P slurry, 94$\%$ (by weight) of Super-P (active material) and 6$\%$ (by weight) of the binder were mixed together in the solvent. The slurries were then doctor-bladed onto molybdenum foil, which was used as the current collector (thickness 0.1 mm, MTI Corp.). The coated sheets were dried in a vacuum oven at 120$^{\circ}$C for 12 hours to adhere the slurry on the substrate and fully evaporate the solvent. The specific loading of the slurry material ranged from 11--12 mg cm$^{-2}$ on each cathode. 

\subsection{Electrolyte preparation}
Anhydrous \ce{AlCl3} (Sigma-Aldrich) and EMImCl (97$\%$, Sigma-Aldrich) were mixed in a molar ratio of 1.3:1, at room temperature. EMImCl was baked in vacuum for 24 hours at 100$^{\circ}$C to remove residual moisture. Small aliquots of AlCl$_3$ was added to EMImCl after every few minutes. The ionic liquid was stirred for 2-3 hours until a clear brown liquid was obtained. Since the electrolyte is hygroscopic in nature, it was prepared in a N$_2$-filled glove box with <0.1 ppm H$_2$O/ O$_2$. 

\subsection{Cell assembly}
PEEK (polyether ether ketone) cells were used for electrochemical measurements. Molybdenum rods were used as current collectors. It was seen previously that steel rods reacted with the electrolyte forming a green-colored substance on the cathode. The slurry coated on molybdenum foil was used as the cathode and placed at bottom of the cell. Two glass microfibers (Grade GF/D, Whatman) were used as separators. 80$\mu$l of the electrolyte was used to wet the separator. Al foil (thickness 0.1 mm, 99$\%$, GoodFellow) used as an anode and placed on top of the separator. It was assembled in a N$_2$-filled glove box. The cell was then sealed and wrapped with a paraffin film to avoid any air or moisture contact. Since this was a two-electrode setup, Al foil was used as both counter and reference electrode. The cell was taken out of the glove box and electrochemical measurements were performed.

\section{Acknowledgement}
We thank Panya Thanwisai and Dr.\ Nonglak Meethong from Khong Kaen University, Thailand for their help with the synthesis of AC from human hair. 

\bibliographystyle{unsrt}  
\bibliography{Thesis2} 
\clearpage
\newpage


\section*{Supporting Information}

\begin{center}
  \textbf{\Huge High voltage carbon-based cathodes for non-aqueous aluminium-ion batteries}\vspace{0.5cm}
  
  \textit{Shalini Divya, Thomas Nann*}
\end{center}

The production of AC consists of carbonisation of a precursor at a temperature below 900$^{\circ}$ C in an inert atmosphere and chemical or physical activation of the carbonised precursor. Activating agents play an important role in determining the porosity of AC \cite{arenas_effect_2004}. Using alkali hydroxides at high temperature creates micropores, which increase the surface area of the material \cite{dong_commercial_2019, liu_hair-based_2017}. In this work, sodium hydroxide (NaOH) was used as the activating agent. The following reaction is taking place within the carbon matrix after addition of NaOH \cite{satish_macroporous_2015}:
\begin{center}
    4NaOH + C $\longrightarrow$ 4Na + 4\ce{CO2} + 2\ce{H2O}
\end{center}

\begin{figure}[h!]
  \centering
  \includegraphics[width=\textwidth]{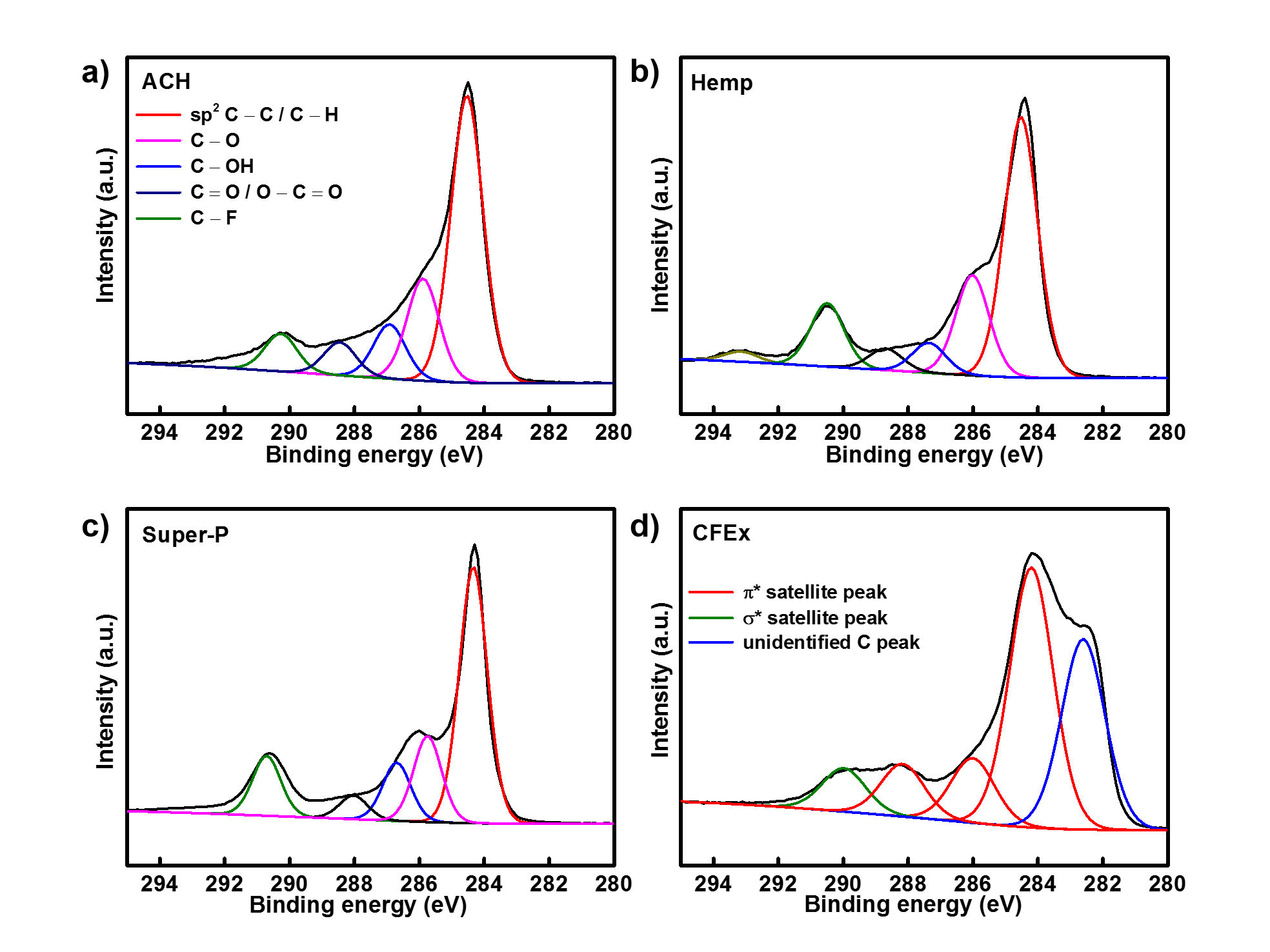}
    \caption{Carbon 1s XPS spectra of pristine a) ACH, b) hemp fibers, c) Super-P and d) CFEx cathodes. While AC from human hair (ACH), hemp fibers and Super-P contain carbonyl functional groups, CFEx cathodes have symmetrical looking $\pi$* and $\sigma$* satellite peaks.}
  \label{fig:XPSC}
\end{figure}

\begin{figure}
  \centering
  \includegraphics[width=0.8\textwidth]{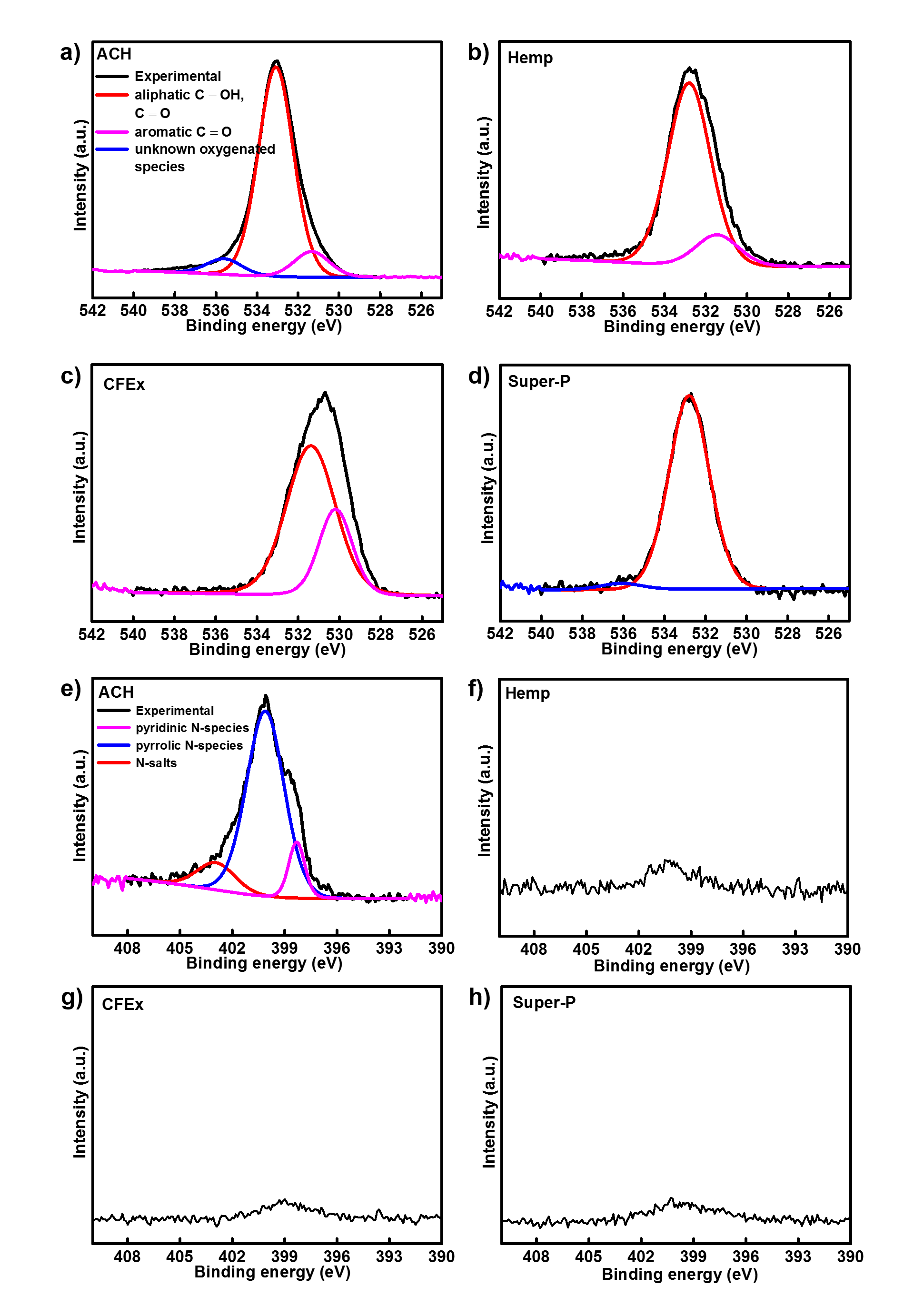}
    \caption{XPS spectra of O 1s orbital for a) ACH, b) hemp fibers c) CFEx and d) Super-P cathodes. Hair and hemp fibers contained significant amounts of aliphatic (red) and aromatic (pink) C=O groups  compared to CFEx and Super-P. Binding energies for N 1s orbital of e) hair, f) hemp fibers g) CFEx and h) Super-P cathodes. Human hair displayed distinct binding energies for pyridinic and pyrrolic N-species; hemp fibers, CFEx and Super-P had smaller amounts of surface proteins.}
  \label{fig:XPSON}
\end{figure}

\begin{figure}[th!]
\centering
\includegraphics[width=0.8\textwidth]{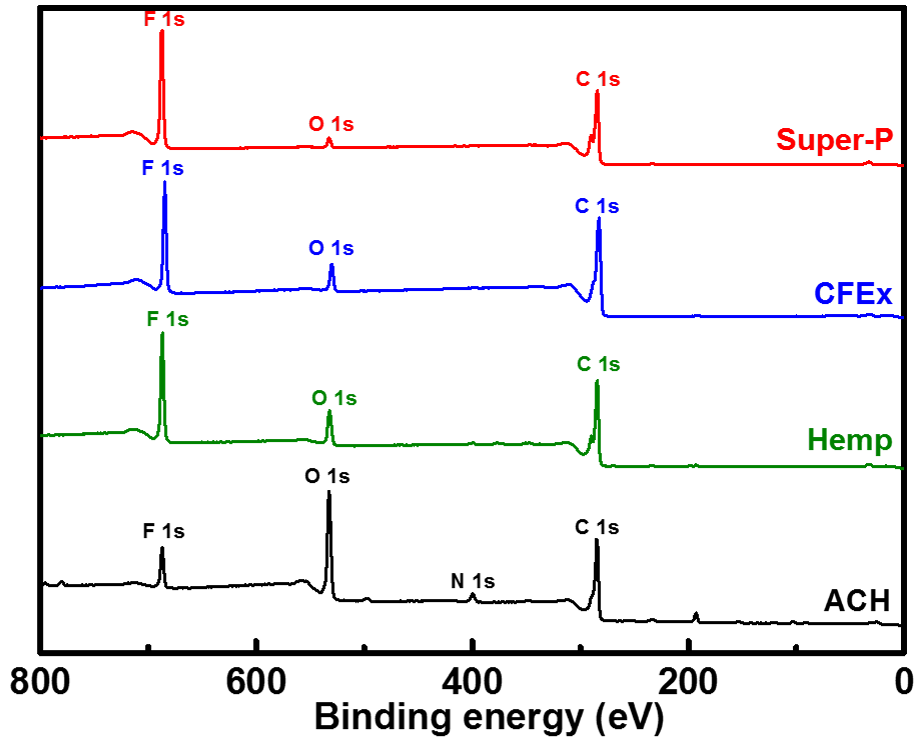}
\caption{Overall spectra of ACH (black), hemp fibers (blue), CFEx (green) and Super-P(red).}
\label{fig:XPSoverall}
\end{figure}

To determine the different bonds and environments existing in all the carbon-based materials, XPS spectra of carbon 1s orbital of all pristine cathodes is shown in Figure \ref{fig:XPSC}. Hair, hemp fibers and Super-P had similar looking peaks for the 1s orbital. Since hair is mainly composed of a protein called keratin, it can be deduced that multiple chemical environments of carbon, shown in Figure \ref{fig:XPSC}a, are derived from Keratin. Sulphide bonds are an essential part of this protein and a carbon-sulphur (C-S) binding energy was observed at 286.94 eV (blue peak) \cite{qian_human_2013}. Spectra for the C 1s orbital of CFEx was uniquely different than the rest due to the presence of several highly symmetrical peaks. The presence of $\pi$ electrons on its surface resulted in  multiple $\pi$ satellite peaks, which are typical in a \ce{C60} molecule \cite{skryleva_xps_2016}. These peaks appear in both high (in green) and low energy ranges (in red) \cite{erbahar_spectromicroscopy_2016, poirier_carbon_1993}. Table \ref{table1} summarises all the functional groups with their respective binding energies for all the pristine cathode materials. Figure \ref{fig:XPSoverall} displays the overall spectra of all tested cathodes. A perfect graphite surface containing only carbon atoms, without heteroatoms, would give a very well-ordered structure. Super-P is produced by partial oxidation of petrochemical precursors \cite{gnanamuthu_electrochemical_2011}. The presence of impurities (such as carbonyl groups) creates defects, resulting in a less graphitic and more amorphous structure \cite{hao_carbonaceous_2013}. The peaks at 288.0 eV in Super-P confirm this observation. If we recall, the presence of these defects was also observed in the form of D-bands in its Raman spectra (Figure \ref{fig:raman}). As reported by Lin and his group in 2011, functional groups that contain oxygen, such as carbonyl and ester groups, improve the wettability of a material \cite{lin_superior_2011}. This increases the availability of the active surface area as more electrolyte ions can interact with the material's surface \cite{younesi_analysis_2015}. Figure \ref{fig:XPSON}a-d, shows various binding energies for the oxygen 1s orbital. In addition to enhancing the wettability of a material \cite{li_effect_2011, oh_oxygen_2014}, oxygen-containing functional groups react with the electrolyte ions and provide pseudo-capacitance \cite{bleda-martinez_role_2005}. Due to the enhanced wettability, the exposure of the chloroaluminate ions was higher in the hair batteries, which resulted in their remarkable performance.
\begin{table}
\caption{Carbon 1s peaks for various binding energies.} \label{table2}
\begin{tabular}{|cccccccc|}
\hline
Active & C-H/ & C-O/ & C=O/ & C-F & Pyrrolic N/ & aliphatic C-O & aromatic C=O\\
material & C-C & C-OH & O-C=O & & Pyridinic N & & \\
\hline
Human hair & 284.5 eV & 285.8 eV & 288.4 eV & 290.2 eV & 400.2 eV/ & 533.0 eV & 531.2 eV\\
& & & & & 398.3 eV & & \\
Hemp fibers & 284.5 eV & 286.0 eV & 288.7 eV & 290.5 eV & 400.3 eV & 532.9 eV & 531.4 eV\\
CFEx & 284.2 eV & 286.0 eV & 288.2 eV & 290.0 eV & 399.3 eV & 531.3 eV & 530.2 eV\\
Super-P & 284.3 eV & 286.7 eV & 288.0 eV & 290.7 eV & 400.2 eV & 532.8 eV & ---\\
\hline
\end{tabular}
\end{table}

\end{document}